\documentclass[12pt]{article}

\usepackage{amssymb}
\usepackage{amsmath}
\usepackage{latexsym}

\catcode`@=11
\renewcommand{\section}{\@startsection{section}{1}{0pt}{\medskipamount}
{\medskipamount}{\large\bf}}
\numberwithin{equation}{section}
\catcode`@=12

\def\beq{\begin{eqnarray}}    
\def\eeq{\end{eqnarray}}      

\def\ln{\,\mbox{ln}\,}                  
\def\sDet{\,\mbox{sDet}\,}              

\def\pa{\partial}                       

\def\={\ =\ }

\begin{document}

\begin{center}

{\Large\bf A systematic study of finite BRST-BV transformations in
field-antifield formalism}

\vspace{18mm}

{\Large Igor A. Batalin$^{(a)}\footnote{E-mail:
batalin@lpi.ru}$\;,
Peter M. Lavrov$^{(b, c)}\footnote{E-mail:
lavrov@tspu.edu.ru}$\;,
Igor V. Tyutin$^{(a)}\footnote{E-mail: tyutin@lpi.ru}$
}

\vspace{8mm}

\noindent ${{}^{(a)}}$
{\em P.N. Lebedev Physical Institute,\\
Leninsky Prospect \ 53, 119 991 Moscow, Russia}

\noindent  ${{}^{(b)}}
${\em
Tomsk State Pedagogical University,\\
Kievskaya St.\ 60, 634061 Tomsk, Russia}

\noindent  ${{}^{(c)}}
${\em
National Research Tomsk State  University,\\
Lenin Av.\ 36, 634050 Tomsk, Russia}

\vspace{20mm}

\begin{abstract}
\noindent We study systematically finite BRST- BV transformations in
the field-antifield formalism. We present explicitly their Jacobians
and the form of a solution to the compensation equation determining
the functional field dependence of finite Fermionic parameters,
necessary to generate arbitrary finite change of gauge-fixing
functions in the path integral.
\end{abstract}

\end{center}

\vfill



\section{Introduction}

It is well known that BRST symmetry \cite{brs1,brs2,t} discovered first
for non-abelian fields
within Faddeev-Popov method \cite{FP} is the fundamental principle
in modern quantizations of arbitrary gauge systems in both Hamiltonian and
Lagrangian formalisms. Parameters of that symmetry are constant
Fermions although they are allowed to be functionals of fields.
Usually, the symmetry is introduced infinitesimally, which means
that its Fermionic  parameters are considered formally as
infinitely-small  quantities. Usual strategy is to show  that the
Jacobian of BRST  transformation does generate arbitrary
variation of gauge-fixing functions in the path integral. This can
be done by choosing necessary functional dependence of BRST
parameters on fields.

The idea to generalize  the BRST symmetry for finite Fermionic
parameters appears quite natural.  Historically, there were several
authors (see \cite{FV,BVhf,BV,BV1,FF,VLT,LT,BF,H,HT,BBD} and
references herein) who worked sporadically with finite BRST
transformations. But the final results were formulated
infinitesimally even in these special cases. In  our recent article
\cite{BLT-FFDT},  we have proposed a systematic study of finite BRST
(with using abbreviation BRST-BFV) transformations in the
generalized Hamiltonian (BFV) formalism  \cite{FV,BVhf,FF}. We have
developed a unique consistent approach to that matter. Its new
strategy was to show that the Jacobian of these finite BRST-BFV
transformations does generate arbitrary finite change of
gauge-fixing functions in the path integral.  In order to do that,
we have formulated the so-called compensation equation determining
the necessary functional field dependence for finite Fermionic
parameters.

In the present paper we will extend our investigations of finite
BRST transformations in Lagrangian (covariant) formalism. The
covariant quantization of gauge theories has made a long way
starting from the famous work of Feynman \cite{F} where
S-matrix non-unitarity in Yang-Mills theories and also in Einstein gravity
within naive quantization rules has been found.Later the emerged problem
was solved by Faddeev and Popov \cite{FP}, and
DeWitt \cite{DeW}. Many authors have contributed to developing
the methods of covariant quantization, as well as to providing them with
various applications. More references can be found in books \cite{HT,W}.
Discovery of supergravity theories
\cite{FvNF,DZ,FvN} and covariant quantization of antisymmetric tensor fields
involved new types of gauge models for which gauge transformations do not form a
gauge group \cite{dWvH,FT}. Direct application of the Faddeev-Popov rules \cite{FP}
leads in the case of these theories to an incorrect result connecting with the
non-unitarity of   physical $S$-matrix. The reason lies in
the structure of gauge transformations for these theories. In this case,
the arising structure coefficients may depend on the fields
of the initial theory, and the gauge algebra of these
transformations may be opened by terms proportional to the equations
of motion. Moreover, attempts of covariant quantization of  gauge
theories with linearly-dependent generators of gauge transformations
result in the understanding of the fact that it is impossible to use
the Faddeev-Popov rules to construct a suitable quantum theory
\cite{Town,HKO}. In turn consistent quantization of supergravity
\cite{K,Nie1} required involving new types of ghosts
(known now as Nielsen-Kallosh ghosts). Therefore, the covariant quantization
of gauge theories demanded in general taking into account many new aspects
such as open algebras, reducible generators and so on as well as
using new approaches. A unique closed approach to the problem of covariant quantization
summarized all these features and attempts was proposed by Batalin and
Vilkovisky \cite{BV,BV1}. The Batalin-Vilkovisky (BV) or field-antifield
formalism gives the rules for the quantization of  general gauge
theories. In the field-antifield formalism, there is a number of rather  specific
features caused by the nontrivial aspects of the antisymplectic
geometry \cite{Schw1}.  The coexistence/interaction between the odd antibracket
and the odd Laplacian should be mentioned first of all in that case.

In the present article, we will develop a unique consistent approach
based on the use of  BRST symmetry with finite Fermionic
parameters,  in the framework of the field - antifield formalism \cite{BV,BV1}.
We will refer to these transformations as finite BRST-BV transformations.
In principle, our new construction does follow the same general
logic as we did in our previous article \cite{BLT-FFDT}.

Finally, let us note that for Yang-Mills theories within the Faddeev-Popov method
\cite{FP}, an attempt to study of finite BRST transformations was undertaken in
\cite{JM} where a differential equation for the Jacobian of such change of variables
in vacuum functional has been proposed, but a solution to this equation
has not been found.  Recently \cite{LL}
it was proved that the problem of finding an explicit form of the Jacobian in Yang-Mills
theories is pure algebraic and can be solved in terms of the BRST variation
of field-dependent parameter. Any finite BRST transformation  of variables
in the generating functional of Green functions is related to modification of
gauge fixing functional \cite{LL,LL1}.
\\

\section{Finite  BRST-BV  Transformations and  Their  Jacobians}

Let
\beq
\label{E2.1}
z^{\alpha}  = \{ \Phi^{A} ; \Phi^*_{A} \},                     
\eeq
be a set of Darboux coordinates of field-antifield phase space,
whose Grassmann parities are
\beq
\label{E2.2}
\varepsilon_{A} = \varepsilon( \Phi^{A} ),\qquad
\varepsilon( \Phi^*_{A} ) = \varepsilon_{A} + 1.               
\eeq
Every anticanonical pair in (\ref{E2.1}) consists of field $\Phi^{A}$ and
antifield $\Phi^*_{A}$, so that the statistics of antifield is opposite
to that of field, in accordance with (\ref{E2.2}). In what follows below,
we will mean the set (\ref{E2.1}) in the sense of condensed DeWitt's
notations, being the capital indices like $\{A\}$ the corresponding
condensed  indices of fields and antifields.

In terms of (\ref{E2.1}), the path integral for the partition function reads
\beq
\label{E2.3}
Z_{\Psi}  = \int D\Phi D\Phi^* D\lambda \exp\left\{ \frac{i}{\hbar} W_{\Psi}\right \},   
\eeq
where the gauge-fixed quantum master action is defined by
\beq
\label{E2.4}
W_{\Psi}  =  W( \Phi, \Phi^* ) + G_{A} \lambda^{A},                            
\eeq
with
\beq
\label{E2.5}
G_{A}  =  \Phi^*_{A} - \Psi(\Phi )\frac{ \overleftarrow{\pa}}{\pa\Phi^{A}}  , 
\eeq
being just a  gauge condition eliminating the antifields in terms of
a field dependent gauge-fixing Fermion $\Psi( \Phi )$. The dynamical
quantum master action $W( \Phi, \Phi^*)$ is defined by the quantum
master equation
\beq
\label{E2.6}
\Delta \exp\left\{ \frac{ i}{ \hbar } W \right\} = 0,                   
\eeq
where
\beq
\label{E2.7}
\Delta = (-1)^{\varepsilon_{A}}  \frac{\pa}{\pa\Phi^{A}}
\frac{\pa}{\pa\Phi^*_{A}} , 
\eeq
is a nilpotent odd Laplacian operator
\beq
\label{E2.8}
\varepsilon( \Delta )=1,\qquad \Delta^{2}=\frac{1}{2} [ \Delta, \Delta ] = 0. 
\eeq
In (\ref{E2.3}), we have also integrated over the Lagrange multipliers
$\lambda^{A}$ with Grassmann parity $\varepsilon( \lambda^{A} ) =
\varepsilon_{A} + 1$. That integration just generates in (\ref{E2.3}) the
gauge-fixing $\delta$-function $\delta( G )$ of (\ref{E2.5}).

The quantum master equation is rewritten in its quadratic form
convenient for $\hbar$-expansion,
\beq
\label{E2.9}
\frac{1}{2} ( W, W )  =  i  \hbar \Delta W,                                          
\eeq
where on the left-hand side  we have used the so-called
antibracket,
\beq
\label{E2.10} ( F, G ) =  F \left( \frac{
\overleftarrow{\pa}}{ \pa\Phi^{A}} \frac{ \overrightarrow{\pa}}{
\pa\Phi^*_{A}} - \frac{\overleftarrow{\pa}}{ \pa\Phi^*_{A}}
\frac{\overrightarrow{\pa}}{\pa\Phi^{A}} \right) G  =
 - ( G, F ) (-1)^{ ( \varepsilon(F) + 1 ) ( \varepsilon(G) + 1 ) }.                  
\eeq
The antibracket (\ref{E2.10}) is odd,
\beq
\label{E2.11}
\varepsilon( ( F, G ) ) = \varepsilon( F ) + \varepsilon ( G ) + 1,                      
\eeq
it  satisfies the Leibnitz rule,
\beq
\label{E2.12}
( F, GH ) = ( F, G ) H + G ( F, H ) (-1)^{ ( \varepsilon(F) +1) \varepsilon(G) }, 
\eeq
it  satisfies the Jacobi identity,
\beq
\label{E2.13}
((F, G),H)(-1)^{(\varepsilon(F)+1)(\varepsilon(H)+1)}+\hbox{cycle}(F,G,H)=0,
\eeq
and it is differentiated by the $\Delta$ operator (\ref{E2.7}),
\beq
\label{E2.14}
\Delta ( F, G ) = ( \Delta F, G ) - ( F, \Delta G ) (-1)^{\varepsilon(F)}.     
\eeq
There exists also the well-known Witten formula,
\beq
\label{E2.15}
\Delta ( F G )  = ( \Delta F ) G + F ( \Delta G ) (-1)^{\varepsilon(F)} +
 ( F, G ) (-1)^{\varepsilon(F)},                                               
\eeq
where the $( F G )$ on the left-hand side means the ordinary
product, not the antibracket.

In the simplest particular case we have the fundamental antibracket
of the anticanonical form,
\beq
\label{E2.16}
( \Phi^{A}, \Phi^{B} ) = 0,\quad   ( \Phi^*_{A}, \Phi^*_{B} ) = 0,
\quad( \Phi^{A},  \Phi^*_{B} ) = \delta^{A}_{B}.                               
\eeq
In terms of the antibracket (\ref{E2.10}), the gauge-fixing functions
$G_{A}$ do commute among themselves,
\beq
\label{E2.17}
( G_{A}, G_{B} ) = 0,                                                          
\eeq
so that the condition $G_{A} = 0$ specifies a Lagrangian hyper-surface
in the field-antifield phase space.

Now, let us define finite field dependent BRST-BV transformations\footnote{
Notice that the transformations (\ref{E2.18}), (\ref{E2.19}) for the fields
$\Phi$ and $\Phi^*$ are really anticanonical for $\mu=\hbox{const}$ and
$\lambda^A$ on the ``mass-shell'' $\pa W_\Psi/\pa\Phi^*_A=0$ $\Rightarrow$
$\lambda^A=-\pa W/\pa\Phi^*_A$.},
\beq
\label{E2.18}
&&\overline{\Phi}^{A} = \Phi^{A} + \lambda^{A} \mu, \\
\label{E2.19}
&&\overline{\Phi^*}_A=\Phi^*_A+
\mu\left(W\frac{\overleftarrow{\pa}}{\pa\Phi^A}\right),\\  
\label{E2.20}
&&\overline{\lambda}^{A} = \lambda^{A}, \\                                     
\label{E2.21}
&&\mu=\mu(\Phi, \lambda ),\quad    \varepsilon( \mu ) = 1.                           
\eeq
Thus, finite Fermionic parameter $\mu$ is only allowed to depend on
fields $\Phi^{A}$ and dynamically-passive Lagrange multipliers
$\lambda^{A}$.

It follows from (\ref{E2.18}) - (\ref{E2.20}) that the transformed gauge-fixed
quantum master action (\ref{E2.4}) has the form
\beq
\nonumber
&&\overline{W_\Psi}=\left.W_\Psi\right|_{z\rightarrow\overline{z}}=W_\Psi+
W\frac{\overleftarrow{\pa}}{\pa\Phi^{A}}
\lambda^A\mu+\mu\left(W\frac{\overleftarrow{\pa}}{\pa\Phi^{A}}\right)
\frac{\pa}{\pa\Phi^*_{A}}W+
 \\
\label{E2.22}
&&+\mu\left(W\frac{\overleftarrow{\pa}}{\pa\Phi^{A}}\right)\lambda^{A}-
\Psi\left(\frac{\overleftarrow{\pa}}{\pa\Phi^{A}}\lambda^{A}\right )^{2}\mu
= W_{\Psi}-\frac{1}{2}(W,W)\mu.                                              
\eeq In (\ref{E2.22}), on the right-hand side of the second
equality, the second term cancels the fourth one, while the fifth
term is zero by itself due to the nilpotency of the operator
squared.

The transformations (\ref{E2.18}), (\ref{E2.19}) cause the following set of
elements of the Jacobi matrix,
\beq
\label{E2.23}
&&\overline{\Phi}^{A}\frac{\overleftarrow{\pa}}{\pa\Phi^{B}}=\delta^{A}_{B}+
\lambda^{A} \mu \frac{ \overleftarrow{\pa}}{ \pa\Phi^{B}} , \\                  
\label{E2.24}
&&\overline{\Phi}^{A} \frac{\overleftarrow{\pa}}{\pa\Phi^*_{B}} = 0, \\                 
\label{E2.25}
&&\overline{\Phi^*}_A\frac{\overleftarrow{\pa}}{\pa\Phi^B}=
\left[\mu\left(W\frac{\overleftarrow{\pa}}{\pa \Phi^{A}}\right)\right]
\frac{ \overleftarrow{\pa}}{\pa\Phi^{B}}, \\
\label{E2.26}
 &&\overline{\Phi^*}_A\frac{\overleftarrow{\pa}}{\pa\Phi^*_B}=\delta_A^B+
\mu \left( W \frac{\overleftarrow{\pa}}{\pa\Phi^{A}}
\frac{\overleftarrow{\pa}}{\pa\Phi^*_{B}}\right).            
\eeq
Due to (\ref{E2.24}),  the complete Jacobian J of the transformation of the
variables (\ref{E2.1}),
\beq
\label{E2.27}
J  = \sDet\left\{ \bar{z}^{\alpha} \frac{\overleftarrow{\pa}}{\pa z^{\beta}} \right\},
\eeq
factorizes to the product of Jacobians of the blocks (\ref{E2.23}) and
(\ref{E2.26}).  In this way, we have
\beq
\label{E2.28}
J = J_{\Phi} J_{\Phi^*},                                                     
\eeq
where
\beq
\label{E2.29}
&&J_\Phi=\sDet\left\{\overline{\Phi}^A\frac{\overleftarrow{\pa}}{\pa\Phi^B}\right\}
=\sDet\left\{\delta^A_B+\lambda^A
\left(\mu\frac{\overleftarrow{\pa}}{\pa\Phi^B}\right)\right\},
 \\
\label{E2.30}
&&J_{\Phi^*}=\sDet\left\{\overline{\Phi^*}_A
\frac{\overleftarrow{\pa}}{\pa\Phi^*_B}\right\}=1-(\Delta W)\mu.                                         
\eeq
It follows from (\ref{E2.22}), (\ref{E2.30}) that
\beq
\nonumber
&&\exp\left\{\frac{i}{\hbar}\overline{W_\Psi}\right\}J_{\Phi^*}=
\exp\left\{\frac{i}{\hbar}W_{\Psi}\right\}
\left[1-\frac{i}{\hbar}\left(\frac{1}{2}( W,W)-i\hbar\Delta W\right)\mu\right]=\\
\label{E2.31}
 &&=\exp\left\{\frac{i}{\hbar}W_{\Psi}\right\},                                                
\eeq
where we have used the quantum master equation (\ref{E2.9}) in the last
equality.

It remains to calculate the factor (\ref{E2.29}). We have
\beq
\label{E2.32}
\ln J_{\Phi} = \int_{0}^{1} d\beta G ^{A}_{B}( \beta ) \lambda^{B}
\left( \mu \frac{\overleftarrow{\pa}}{\pa\Phi^{A}}\right) (-1)^{ \epsilon_{A} },                
\eeq
where $G^{A}_{B}( \beta )$ is defined by the equation
\beq
\label{E2.33}
\left[ \delta^{A}_{B} + \beta \lambda^{A} \left( \mu \frac{\overleftarrow{\pa}}{\pa\Phi^{B}}
\right ) \right] G^{B}_{C} ( \beta) =
\delta^{A}_{C}.                                                                 
\eeq
It follows immediately from (\ref{E2.33}) that
\beq
\label{E2.34}
G^{A}_{B} = \delta^{A}_{B} -  \beta \lambda^{A}
( 1 + \beta \kappa )^{-1} \left( \mu \frac{\overleftarrow{\pa}}{ \pa\Phi^{B}}\right ),         
\eeq
where the functional $\kappa$ equals
\beq
\label{E2.35}
\kappa  = \mu \frac{\overleftarrow{\pa}}{\pa\Phi^{A}}\lambda^{A}.           
\eeq
By substituting (\ref{E2.34}) into (\ref{E2.32}) we get the
following $\beta$-integral
\beq \label{E2.36} \ln J_{\Phi} =
\int_{0}^{1} d\beta [ - \kappa + \beta ( 1 + \beta \kappa )^{-1}
\kappa^{2} ] = - \int_{0}^{1} d\beta \kappa ( 1 + \beta \kappa
)^{-1} =
- \ln( 1 + \kappa ),                                                             
\eeq
so that
\beq
\label{E2.37}
J_{\Phi}  =  ( 1 + \kappa )^{-1}.                                            
\eeq

\section{Compensation   Equation and  Its  Explicit  Solution}

Now, we would like to use the field Jacobian (\ref{E2.37}) to generate arbitrary
finite change $\delta \Psi$ of the gauge Fermion $\Psi$  in the path integral
(\ref{E2.3}) with the action (\ref{E2.4}), (\ref{E2.5}),
\beq
\label{E3.1}
\Psi \rightarrow \Psi_{1} =  \Psi + \delta \Psi.                              
\eeq
Let us proceed with the path integral in the new variables
(\ref{E2.18}) - (\ref{E2.20}),
\beq \label{E3.2} Z_{\Psi}=\int
D\overline{\Phi}D\overline{\Phi^*}D\overline{\lambda}
\exp\left\{\frac{i}{\hbar}\overline{W_\Psi}\right\}=
\int D\Phi D\Phi^* D\lambda J_{\Phi}\exp\left\{\frac{i}{\hbar}W_{\Psi}\right\},         
\eeq
where we have used  (\ref{E2.28}),  (\ref{E2.31}).
In order to provide for the change (\ref{E3.1}), let us
require the condition
\beq
\label{E3.3}
J_\Phi=\exp\left\{\!\!-\frac{i}{\hbar}\left(\delta\Psi
\frac{\overleftarrow{\pa}}{\pa\Phi^A}\lambda^{A}\right)\right\}                                                                    
\eeq
to  hold. Then we arrive at
\beq
\label{E3.4}
Z_{\Psi_{1}} = Z_{\Psi},                                                        
\eeq
which means the $\Psi$-independence for the partition function.

It is also true that the quantum mean value $< \mathcal{O} >_{\Psi}$
with the weight functional $\exp\{(i/\hbar)W_{\Psi} \}$ does
not depend on $\Psi$ for any physical observable $\mathcal{O}$ annihilated by
the Fermionic nilpotent $\sigma=(W,...)-i\hbar\Delta(...)$,
$\sigma\mathcal{O} = 0$. The main idea of the proof is that the product
$\mathcal{O}\exp\{(i/\hbar)W\}$ is annihilated by the operator $\Delta$ due
to (\ref{E2.6}) and the definition of $\sigma$, so that
$\overline{W}=W+(\hbar/i)\ln\mathcal{O}$ satisfies eq. (\ref{E2.6}) as well. Then
we refer to the usual argument with $\overline{W}$ standing for $W$.

Due to (\ref{E2.35}), (\ref{E2.37}),  the condition (\ref{E3.3})
is rewritten  in the form
\beq
\label{E3.5}
 \mu \frac{\overleftarrow{\pa}}{\pa\Phi^{A}} \lambda^{A}=
\exp\left\{\frac{i}{\hbar}\left(\delta\Psi\frac{\overleftarrow{\pa}}{\pa\Phi^{A}}
\lambda^{A}\right)\right\}-1.   
\eeq
We call the condition (\ref{E3.3}) or (\ref{E3.5}) "a compensation equation".
Actually, that equation
has to determine the necessary field dependence for $\mu$.

There exists an obvious explicit solution to the compensation equation (\ref{E3.5}),
\beq
\label{E3.6}
\mu=\mu(\delta\Psi)=\mu(\Phi,\lambda;\delta\Psi)=\frac{i}{\hbar}f(x)\delta\Psi,                                             
\eeq
where the functional $x$ equals
\beq
\label{E3.7}
x=\frac{i}{\hbar}\delta\Psi\frac{\overleftarrow{\pa}}{\pa\Phi^{A}}\lambda^{A},
\eeq
and
\beq
\label{E3.8}
f(x)=(\exp(x)-1)\,x^{-1}.                                    
\eeq As the operator on the right-hand side in (\ref{E3.7}) is
nilpotent, the latter does annihilate the $x$ (\ref{E3.7}) and any
function of it.  As the same nilpotent operator stands on the
left-hand side in (\ref{E3.5}),  it applies nontrivially only to the
rightmost factor $\delta\Psi$ in (\ref{E3.6}), which results exactly
in having the factor $x^{-1}$ in (\ref{E3.8}) canceled. In this way,
we have confirmed immediately the equation (3.5) to hold.

Notice,  that for finite change $\delta\Psi$ the solution (\ref{E3.6}) is in general
$\lambda$ dependent.
However, in the first order in $\delta\Psi$, explicit solution (\ref{E3.6}) takes the
 usual form
\beq
\label{E3.9}
 \mu(\delta\Psi) =\frac{i}{\hbar}\delta\Psi+O\big((\delta\Psi)^2\big).                                        
\eeq which is $\lambda$ independent as far as the $\delta\Psi$ does
the same.

\section{$\lambda$-Differential as a  BRST-BV Generator for  Fields}

Let us consider the Fermionic nilpotent operator
\beq
\label{E4.1}
&&\overleftarrow{d}  =  \frac{\overleftarrow{\pa}}{\pa\Phi^{A}} \lambda^{A}, \\ 
\label{E4.2}
&&\epsilon( \overleftarrow{d} )  =  1,\qquad
( \overleftarrow{d} )^{2}  =  \frac{1}{2} [ \overleftarrow{d}, \overleftarrow{d}]  =  0.  
\eeq
In terms of (\ref{E4.1}) the action (\ref{E2.4}), (\ref{E2.5}) is rewritten as
\beq
\label{E4.3}
W_{\Psi} = W( \Phi, \Phi^* )+( \Phi^*_{A} \Phi^{A} - \Psi(\Phi)) \overleftarrow{d}.     
\eeq
The transformation (\ref{E2.18}) of fields $\Phi^{A}$ takes the form
\beq
\label{E4.4}
\overline{\Phi}^{A} =  \Phi^{A} ( 1 + \overleftarrow{d} \mu ).                        
\eeq
Thus, the operator (\ref{E4.1})  is a generator for finite  BRST-BV field transformation.

The formula (\ref{E2.37}) for the Jacobian $J_{\Phi}$ is rewritten as
\beq
\label{E4.5}
J_{\Phi} = J_{\Phi}(\mu) =[ 1 + ( \mu \overleftarrow{d} ) ]^{-1}.                    
\eeq
The compensation equation (\ref{E3.5}) takes the form
\beq
\label{E4.6}
\mu\overleftarrow{d} =
\exp\left\{\frac{i}{\hbar}\left(\delta\Psi\overleftarrow{d}\right)\right\}-1.            
\eeq
The $x$ in (\ref{E3.7}) can be represented as
\beq
\label{E4.7}
x=\frac{i}{\hbar}\delta\Psi\overleftarrow{d}.                                
\eeq
Thus, we conclude that all the main objects in our
consideration can be expressed naturally in terms of a single
quantity that is the BRST-BV field generator (\ref{E4.1}), also
called "a $\lambda$-differential".

Notice that the introduced field transformations (\ref{E4.4}) form a group.
Indeed, let us rewrite
(\ref{E4.4}) in the form
\beq
\label{E4.8}
\overline{\Phi}^{A}  = \Phi^{A} \overleftarrow{T}( \mu ), \;
\overleftarrow{T}( \mu ) = 1 + \overleftarrow{d} \mu.            
\eeq
Then the group composition law of the transformations (\ref{E4.8}) reads
\beq
\label{E4.9}
\overleftarrow{T}( \mu_{1} ) \overleftarrow{T}( \mu_{2} ) =
\overleftarrow{T}( \mu_{12} ),           
\eeq
where
\beq
\label{E4.10}
\mu_{12}  = \mu_{1}  + [J_{\phi}(\mu_{1})]^{-1} \mu_{2},                
\eeq
where $J_{\Phi}(\mu_{1})$ is the Jacobian (\ref{E4.5}) with $\mu_{1}$ standing for $\mu$.
Indeed, due to the nilpotency (\ref{E4.2}) of $\overleftarrow{d}$, we have
\beq
\label{E4.11}
\overleftarrow{T}( \mu_{1} ) \overleftarrow{T}( \mu_{2} ) =
1 +  \overleftarrow{d} \mu_{1}  + \overleftarrow{d} \mu_{2}  +
\overleftarrow{d}\mu_{1} \overleftarrow{d}\mu_{2} =
1 +  \overleftarrow{d} \mu_{1}  + \overleftarrow{d} \mu_{2}  +
\overleftarrow{d}( \mu_{1} \overleftarrow{d} ) \mu_{2}.              
\eeq
By inserting here
\beq
\label{E4.12}
( \mu_{1} \overleftarrow{d} ) =  [ J_{\phi}( \mu_{1} ) ]^{-1} - 1.                 
\eeq
we arrive at (\ref{E4.10}).

Moreover, it follows from (\ref{E4.9}) that the algebra of the group generators
has the form
\beq
\label{E4.13}
[ \overleftarrow{d} \mu_{1}, \overleftarrow{d} \mu_{2} ] =
\overleftarrow{d} \mu_{[12]},           
\eeq
where
\beq
\label{E4.14}
\mu_{[12]}  = \mu_{12} - \mu_{21}  = - ( \mu_{1} \mu_2 ) \overleftarrow{d}.     
\eeq

\section{Ward Identities  Dependent  of Finite  BRST-BV \\Parameters/Functionals}

As we have defined finite BRST-BV transformations, it appears quite natural to use them
immediately to derive the corresponding modified version of the Ward identity. We will
do that  just in terms of BRST-BV field generator introduced in Section 4.

As usual for that matter, let us proceed with the external-source dependent generating
functional
\beq
\label{E5.1}
Z_{\Psi}( \zeta ) =
 \int D\Phi D\Phi^* D\lambda \exp\left\{ \frac{ i}{ \hbar } W_{\Psi}( \zeta)\right\} ,       
\eeq
where
\beq
\label{E5.2}
W_{\Psi}( \zeta )  =  W_{\Psi} + \zeta_{A} \Phi^{A},               
\eeq
$\zeta_{A}$ are arbitrary external sources to the fields
$\Phi^{A}, \varepsilon( \zeta_{A} ) = \varepsilon_{A}$. Notice that we do not
introduce their own sources to antifields $\Phi^*_{A}$.
Of course, in the presence of non-zero external source, the path integral
(\ref{E5.1}) by itself is in general actually dependent of gauge Fermion $\Psi$.
However, this dependence has a special form and the equivalence
theorem \cite{KT}, applying in physical sector, makes possible to establish
that the physical quantities do not depend on gauge. In its turn, the Ward
identity measures the deviation of the path integral from being
gauge-independent.

Let  us perform in (\ref{E5.1})  the change (\ref{E2.18})-(\ref{E2.20}) of integration
variables, with arbitrary
finite $\mu( \phi, \lambda )$. Then, by using (\ref{E2.31}) and (\ref{E4.5}),
we get what we call "a modified Ward identity",
\beq
\label{E5.3}
\left\langle \left[ 1 + \frac{i}{ \hbar } \zeta_{A} ( \Phi^{A}\overleftarrow{d} ) \mu\right ]
 [1 +( \mu \overleftarrow{d} ) ]^{-1}\right \rangle_{\Psi, \zeta} = 1,
\eeq                                 
where we have denoted the source dependent mean value
\beq
\label{E5.4}
\langle(...)\rangle_{\Psi,\zeta}=[Z_{\Psi}(\zeta)]^{-1}\int D\Phi D\Phi^* D\lambda
(...)\exp\left\{\frac{i}{\hbar}W_{\Psi}(\zeta)\right\}, \qquad <1>_{\Psi,\zeta}=1,   
\eeq
related to the source dependent action (\ref{E5.2}). By construction,
in (\ref{E5.3}), both $\zeta_{A}$ and $\mu(\Phi,\lambda)$ are arbitrary. The
presence of arbitrary $\mu(\Phi,\lambda)$ in the
integrand in (\ref{E5.3}) reveals the implicit dependence of generating functional
(\ref{E5.1})  on the
 gauge-fixing Fermion $\Psi$ for non-zero external source $\zeta_{A}$.

For a constant $\mu$, $\mu = \hbox{const}$, the latter does drop-out completely, and we get
from (\ref{E5.3})
\beq
\label{E5.5}
\langle \zeta_{A} \lambda^{A} \rangle_{\Psi, \zeta}  =  0.                                  
\eeq
By using the representation
\beq
\label{E5.5a}
\lambda^A \exp\left\{ \frac{i}{ \hbar }G_B\lambda^B\right\}=
\frac{\hbar}{i}\frac{\pa}{\pa\Phi^*_A}\exp\left\{ \frac{i}{ \hbar }G_B\lambda^B\right\},
\eeq
and integrating over the antifields $\Phi^*_A$ by part, (\ref{E5.5}) is rewritten
as
\beq
\label{E5.6}
\left\langle\zeta_A\left(\frac{\pa}{\pa\Phi^*_{A}}W\right)\right\rangle_{\Psi,\zeta}=0,   
\eeq
which is exactly the standard form of a Ward identity in the field-antifield formalism.

By identifying the $\mu$ in (\ref{E5.3})  with the solution  (\ref{E3.6})
to the compensation equation (\ref{E4.6}),
it follows according to our result in Section 3,
\beq
\label{E5.7}
Z_{\Psi_{1}}( \zeta ) = Z_{\Psi}(\zeta)\left [ 1 +
\left\langle \frac{i}{  \hbar } \zeta_{A} ( \Phi^{A}\overleftarrow{d} )
\mu( - \delta\Psi ) \right\rangle_{\Psi, \zeta}\right ].              
\eeq
Formula (\ref{E5.7}) generalizes the gauge independence (\ref{E3.4})
of the partition function to the presence of the external source.

\section{Discussions}

We have introduced  the conception of finite  BRST-BV
transformations in the field-antifield quantization formalism \cite{BV,BV1}
for general gauge-field dynamical systems. It was shown  that the
Jacobian  of finite BRST-BV transformations, being the main
ingredient of the approach, can be calculated explicitly in terms of
the corresponding generator applied to finite field-dependent
functional parameters of these transformations. We have introduced
the compensation equation providing  for a connection between the
generating functionals formulated for a given dynamical system in two
different gauges. We have extended the proof of gauge independence of
partition function and quantum mean values of physical observables as to
the case of finite variations of gauge-fixing functional. We have found an
explicit solution to the
compensation equation proposed. We have studied the algebra and the
group composition law of finite BRST-BV transformations. As a
by-product, we have developed a technique using the so-called
$\lambda$-differential, provided for deriving in a simple way the
Ward identity and connection between the generating functionals
of Green functions written in two different gauges.

In conclusion, we would like to present in short an alternative view on the
role of finite BRST transformations as respected by the path integral
(\ref{E2.3}). Let us rewrite the latter in the form
\beq \label{E6.1}
Z_\Psi=\int D\Phi D\Phi^*D\lambda\exp\left\{\frac{i}{\hbar}(W+X)\right\},    
\eeq
where $W$ satisfies (\ref{E2.6}) while $X$ is given by
\beq
X = G_{A} \lambda^{A}                                         
\eeq
with $G_{A}$ defined in (\ref{E2.5}). Due to (\ref{E2.17}), it follows that
\beq \label{E6.3}
\Delta\exp\left\{\frac{i}{\hbar}X\right\}=0,          
\eeq
which is symmetric to (\ref{E2.6}). In the integrand in (\ref{E6.1}), we have
the BRST-BV symmetry as represented in its infinitesimal form
\beq \label{E6.4}
\delta z^\alpha=(z^\alpha,-W + X)\mu+\frac{\hbar}{i}(z^\alpha,\mu).   
\eeq

Of course, in principle, we could try to reformulate (\ref{E6.4}) at the level
of finite transformations, similarly to what we did with respect to
(\ref{E2.18}), (\ref{E2.19}). Instead of doing that here, let us consider now
finite equivalence transformations acting on the space of solutions to
(\ref{E6.3}),
\beq
\exp\left\{\frac{i}{\hbar}X'\right\}=\exp\{-[F,\Delta]\}\exp\left\{\frac{i}{\hbar}X\right\},         
\eeq
where a function $F$ is a finite Fermionic generator.  Due to the relation
\beq
[F,\Delta]= (\Delta F)-(F, ... ),       
\eeq
we have
\beq \label{E6.7}
X'=\exp\{(F, ...)\}X+i\hbar f((F,...))\Delta F,         
\eeq
where $(F,...)$ means the left adjoint action of the antibracket,
\beq
(F,...)G=(F,G),        
\eeq and the function $f(x)$ is given by (\ref{E3.8}). On the
right-hand side in (\ref{E6.7}), the first term is an anti-canonical
transformation with finite Fermionic generator $F$, while the second
term is a half of a logarithm of the Jacobian of that transformation
up to $(-i\hbar)$. If we choose \beq
F = - \delta \Psi( \Phi )   
\eeq
with finite $\delta\Psi$, then
\beq
X'=G'_A\lambda^A,           
\eeq
where
\beq
G'_A=\Phi^*_A-\Psi'\frac{\overleftarrow{\pa}}{\pa\Phi^A},  
\eeq
and
\beq
\Psi' = \Psi + \delta \Psi.          
\eeq
One can rewrite the formula (\ref{E6.7}) as
\beq
X'-X=-f((F,...))\sigma_X F,
\eeq
where
\beq
\label{E6.14}
\sigma_X=-i\hbar\exp\left\{-\frac{i}{\hbar}\right\}\Delta\exp\left\{\frac{i}{\hbar}\right\}=
(X,...)-i\hbar \Delta,
\eeq
while in the second equality we have used the quantum master equation (\ref{E6.3}).

An infinitesimal form of (\ref{E6.3})
\beq
\delta X=-\sigma_X F+O(F^2),
\eeq
shows that $-\sigma_X$ is a generator to the corresponding variation $\delta X$. As the operator
$\Delta$ is nilpotent, it follows from the first equality in (\ref{E6.14}) that the operator
$\sigma_X$ is nilpotent as well.

Of course, it should be noticed that the equivalence transformation
(\ref{E6.7}) as presented above, was not generated by any change of
integration variables in (\ref{E6.1}). However it can be shown that such a
change of variables can be deduced from (\ref{E6.4}) by "integrating" the
latter to the level of finite transformation.

In the present article, we have explored the "$Sp(1)$-version" of
the BRST symmetry with a single Fermionic parameter, in the
field-antifield formalism. Few years ago, we have proposed the
$Sp(2)$-version of the field-antifield formalism, based on the
conception of the extended BRST symmetry with two Fermionic
parameters \cite{BLTl1,BLTl2,BLTl3}. It seems very interesting to extend
the results obtained above to the case of $Sp(2)$-version of the BRST
symmetry.
\\

\section*{Acknowledgments}
\noindent
 I. A. Batalin would like  to thank Klaus Bering of Masaryk
University for interesting discussions. The work of I. A. Batalin is
supported in part by the RFBR grants 14-01-00489 and 14-02-01171.
 The work of P. M. Lavrov is partially supported by the Ministry of Education and Science of
Russian Federation, grant TSPU-122, by the Presidential grant
88.2014.2 for LRSS and  by the RFBR grants 12-02-00121 and 13-02-91330. The work of
I. V. Tyutin is partially supported by the RFBR grant 14-01-00489.
\\


\begin {thebibliography}{99}
\addtolength{\itemsep}{-8pt}



\bibitem{brs1}
C. Becchi, A. Rouet and R. Stora,
{\it
Phys. Lett. B} {\bf 52}, 344 (1974).

\bibitem{brs2}
C. Becchi, A. Rouet and R. Stora,
{\it
Ann. Phys. (N. Y.)} {\bf  98}, 287  (1976).

\bibitem{t}
I. V. Tyutin,
{\it Gauge invariance in field theory and statistical physics
in operator formalism}, Lebedev Institute preprint  No.  39  (1975),
arXiv:0812.0580[hep-th].


\bibitem{FP}
L. D. Faddeev and V. N. Popov, {\it
Phys. Lett. B} {\bf 25}, 29 (1967).


\bibitem{FV}
E. S Fradkin and G. A. Vilkovisky, {\it  Phys. Lett. B} {\bf 55}, 224 (1975).

\bibitem{BVhf}
I. A. Batalin and G. A. Vilkovisky, {\it Phys. Lett.
B} {\bf 69}, 309 (1977).

\bibitem{BV}
I. A. Batalin and G. A. Vilkovisky, {\it  Phys. Lett. B} {\bf 102}, 27 (1981).

\bibitem{BV1}
I. A. Batalin and G. A. Vilkovisky, {\it  Phys. Rev. D} {\bf 28}, 2567 (1983).

\bibitem{FF}
E. S. Fradkin and T. E. Fradkina, {\it
 Phys. Lett. B} {\bf 72}, 343 (1978).

\bibitem{VLT}
B. L. Voronov, P. M. Lavrov and I. V. Tyutin, {\it
Sov. J. Nucl. Phys.} {\bf 36}, 292 (1982).

\bibitem{LT}
P. M. Lavrov and I. V. Tyutin,
{\it
Sov. Phys. J.} {\bf  25}, 639 (1982).

\bibitem{BF}
I. A. Batalin and E. S. Fradkin,
{\it
Phys. Lett. B} {\bf 122}, 157 (1983).

\bibitem{H}
M. Henneaux,
{\it
Phys. Rep.} {\bf 126}, 1 (1985).

\bibitem{HT}
M. Henneaux and C. Teitelboim,
{\it Quantization of gauge systems} (Princeton University Press, Princeton,1992).

\bibitem{BBD}
I. A. Batalin, K. Bering and P. H. Damgaard, {\it Nucl. Phys. B} {\bf 739}, 389
(2006).

\bibitem{BLT-FFDT}
I. A. Batalin, P. M. Lavrov and I. V. Tyutin,
{\it Int. J. Mod. Phys. A } {\bf 29}, 1450127 (2014).

\bibitem{F}
R. P. Feynman,
{\it
Acta Phys. Pol.} {\bf 24}, 697 (1963).

\bibitem{DeW}
B. S. De Witt,
{\it
Phys. Rev.} {\bf 162} (1967) 1195.

\bibitem{W}
S. Weinberg,
{\it The Quantum theory of fields}, Vol.II
(Cambridge University Press, 1996)


\bibitem{FvNF}
D. Z. Freedman, P. van Nieuwenhuizen and S. Ferrara,
{\it
Phys. Rev. D} {\bf 13}, 3214 (1976).

\bibitem{DZ}
S. Deser and B. Zumino,
{\
Phys. Lett. B} {\bf 62}, 335 (1976).

\bibitem{FvN}
D. Z. Freedman and P. van Nieuwenhuizen,
{\it
Phys. Rev. D} {\bf 14}, 912 (1976).

\bibitem{dWvH}
B. de Wit and J. W. van Holten,
{\it
Phys. Lett. B} {\bf 79}, 389 (1978).

\bibitem{FT}
D. Z. Freedman and P. K. Townsend,
{\it
Nucl. Phys. B} {\bf 177}, 282 (1981).


\bibitem{Town}
P. K. Townsend,
{\it
Phys. Lett. B} {\bf 88}, 97 (1979).


\bibitem{HKO}
H. Hata, T. Kugo and N. Ohta,
{\it
Nucl. Phys. B} {\bf 178}, 527 (1981).

\bibitem{K}
R. E. Kallosh, {\it
Nucl. Phys. B} {\bf 141}, 141 (1978).

\bibitem{Nie1}
N. K. Nielsen,
{\it
Nucl. Phys. B} {\bf 140}, 494 (1978).


\bibitem{Schw1}
A. Schwarz, {\it
Comm. Math. Phys.} {\bf 155}, 249 (1993).



\bibitem{JM}
S. D. Joglekar and B. P. Mandal,
{\it
 Phys.Rev. D} {\bf 51}, 1919 (1995).

\bibitem{LL}
P. M. Lavrov and O. Lechtenfeld,
{\it  Phys. Lett. B} {\bf 725}, 382 (2013).


\bibitem{LL1}
 P. M. Lavrov and O. Lechtenfeld,
{\it Phys. Lett. B} {\bf 725}, 386 (2013).


\bibitem{Witten}
E. Witten, {\it
Mod. Phys. Lett. A} {\bf 5}, 487 (1990).

\bibitem{KT}
R. E.  Kallosh and I.V. Tyutin,
{\it  Sov. J. Nucl. Phys.} {\bf 17}, 98 (1973).

\bibitem{BLTl1}
I. A. Batalin, P. M. Lavrov and I. V. Tyutin,
{\it
J. Math. Phys.} {\bf 31}, 1487 (1990).

\bibitem{BLTl2}
I. A. Batalin, P. M. Lavrov and I. V. Tyutin,
{\it
 J. Math. Phys.} {\bf 32}, 532 (1991).

\bibitem{BLTl3}
I. A. Batalin, P. M. Lavrov and I. V. Tyutin,
{\it
 J. Math. Phys.} {\bf 32}, 2513 (1991).

\end{thebibliography}
\end{document}